\documentclass[twocolumn,showpacs,preprintnumbers]{revtex4}

\usepackage{graphicx,graphics,epsfig,subfigure,times,bm,bbm,amssymb,amsmath,amsfonts,amsthm,mathrsfs,MnSymbol}
\usepackage[matrix,frame,arrow]{xypic}
\usepackage[pdfstartview=FitH]{hyperref}
\usepackage{epstopdf}
\usepackage{pifont}
\hypersetup{
    colorlinks=true,       	
    linkcolor=red,          	
   citecolor=magenta,        
    filecolor=magenta,      	
    urlcolor=cyan,           	
    runcolor=cyan
}
\usepackage[pdftex]{color}
\usepackage{braket}
\usepackage{enumerate}
\usepackage[normalem]{ulem}

\usepackage{subfigure}
\usepackage{overpic}

\usepackage{multirow}
\newcommand{\be}{\begin{equation}}
\newcommand{\ee}{\end{equation}}
\newcommand{\ba}{\begin{eqnarray}}
\newcommand{\ea}{\end{eqnarray}}

\newcommand{\ignore}[1]{}

\begin{document}
\title{Effective dynamics of  cold atoms \\ flowing in two  ring shaped optical potentials with tunable tunneling}
\author{Davit Aghamalyan$^{1}$, Luigi Amico$^{1,2}$, and L.C Kwek$^{1,3}$}
\affiliation{$^{1}$ Centre for Quantum Technologies, National University of
Singapore, 3 Science Drive 2, Singapore 117543}
\affiliation{$^{2}$ CNR-MATIS-IMM $\&$
Dipartimento di Fisica e Astronomia,   Via S. Sofia 64, 95127 Catania, Italy}
\affiliation{$^{3}$National Institute of
Education and Institute of Advanced Studies, Nanyang Technological
University, 1 Nanyang Walk, Singapore 637616}
\date{\today }
\begin{abstract}
We study the current dynamics of  coupled atomic condensates flowing  in two ring-shaped optical potentials.
We provide a  specific setup  where the ring-ring  coupling can be tuned in experimentally feasible way.
It is demonstrated that the imaginary time effective action of the system in a  weak coupling regime provides a two-level-system-dynamics for the phase slip across the two rings. Through  two-mode Gross- Pitaevskii  mean field equations, the real time dynamics of the population imbalance  and the phase difference between of the two condensates   is thoroughly analyzed analytically, as function of  the relevant  physical parameters  of the system. In particular,  we find that the macroscopic quantum self trapping  phenomenon is induced in the system if the flowing currents assume a non vanishing difference.

\end{abstract}

\maketitle
\section{Introduction}
Quantum technology has been leading to  realistic applications. To this end, the physical community have been combining ideas and techniques from many fields, including quantum optics, quantum information and condensed matter physics.  Although  devising new technological applications remains a defining goal in the field,  quantum technologies allow us also  to   explore new physical regimes, disclosing  new fundamental science.
Ultracold atoms loaded into optical lattices has been playing an important role in this context\cite{coldatoms,Bloch, rydberg,Raithel, Friebel, Raizen, Guidoni,  Petsas,Deutsch}: they are precise  and easily accessible quantum simulators \cite{Feynman},  assisting both in the solution of puzzling problems coming from other fields (like solid state physics) and in the engineering of new quantum phases of extended systems; at the same time,  they provide new devices for future technologies, like quantum metrology and quantum computation.
%
%
%

In the scenario depicted above, it is desirable  to work with different spatial optical lattice configuration.  Ring shaped  optical lattices, in particular, allow us to engineer textbook periodic boundary conditions in many body systems, and pave the way for exploiting the currents in the lattices  as 'degrees of freedom' for new quantum devices. Such optical lattices can be generated by employing Laguerre-Gauss laser beams \cite{Luigi,ferris}, by using a rapidly-moving laser beam that "paints" a time averaged optical dipole potential\cite{potential exp}, or by a spatial light modulator (SLM) which imprints a controlled phase onto a collimated
laser beam\cite{Flux qubit}. Light fields of different circular structure has been theoretically proposed \cite{Kolke} by making use of Bessel laser beams.
The currents can be generated in several ways: by rotating Bose condensates \cite{rot1,rot2}, by shining the atoms with electrical fields and making use of so-called synthetic gauge fields \cite{synth1,synth2}, by using conical shaped magnetic field \cite{Luigi}or by imprinting a  Berry-phase\cite{berry}.

Aside from other applications\cite{q-motors}, neutral  currents in ring-shaped optical potentials  are natural candidates to provide a realization of  Josephson junctions flux qubit analog \cite{Mooij,Paladino}.
This would exploit  the best features of the superconducting  flux qubits  together with the typically low decoherence time of the cold atom based qubits. In the Ref. \cite{Flux qubit} it was evidenced that the program can be indeed  realized, constructing the qubit with bosons loaded in single ring lattice interrupted by a weak  link.

In this paper, in contrast, we study a specific device comprising two  {\it homogeneous} ring-shaped potential with ring-ring coupling. In view of the possible 'scalability' of the system, we provide a feasible way to construct a ring-ring interaction, mimicking the inductive coupling in the devices based on charged currents (like  before mentioned SQUID-based devices). Indeed, in our specific  setup, the coupling can be tuned with simple operations. The system is envisaged to be loaded with bosonic atoms, thus realizing a Bose-Hubbard ladder. We demonstrate that the imaginary-time  dynamics of the phase difference across the two weakly coupled rings is controlled by double-well potential. Therefore, the system  of two {homogenous} tunnel-coupled rings indeed defines a qubit.   The real time dynamics is studied within  mean field two-mode Gross-Pitaevskii equations.  For the  analysis, we benefit from Ref. \cite{Kenkre,Smerzi,Smerzi1}.
We demonstrate  that the system is characterized by  Macroscopic Quantum Self-Trapping(MQST)\cite{Milburn,Smerzi,Smerzi1,Albiez}, the atomic  analog  of the solid state polaronic non-linear self-localization  phenomenon  due to the strong electron-lattice interaction\cite{Landau,Kenkre}.  In contrast  to the polaron case, the nonlinearity of the  Bose-Einstein condensate self trapping arises from the  many-particle interactions.

The paper is outlined  as follows. In section \ref{sec:experiment_setup}, we describe the  experimental setup realizing   two homogenous ring-shaped optical lattices with tunable interaction between of them. In section \ref{sec:qubit_dynamics}, we describe how the phase differences along the wells of the two rings can be integrated out in the imaginary time action, this leading to an effective qubit  dynamics for the phase difference across the two rings.   In section \ref{sec:real_dynamics}, we investigate the real-time dynamics for two coupled ring-shaped optical lattice based on two coupled Gross-Pitaevskii equations, and describe the various possible regimes. In section \ref{sec:Allowed regions of oscillation}, regions of oscillations with MQST and phase space diagrams are detailed for the different values of the relevant  physical parameters. Finally, we summarize our results in section \ref{sec:conclusions}.

\section{\protect\normalsize Bosonic atoms loaded in two-rings optical potential with tunable coupling}\label{sec:experiment_setup}

\begin{figure}
\begin{center}
{\includegraphics*[scale = 0.3]{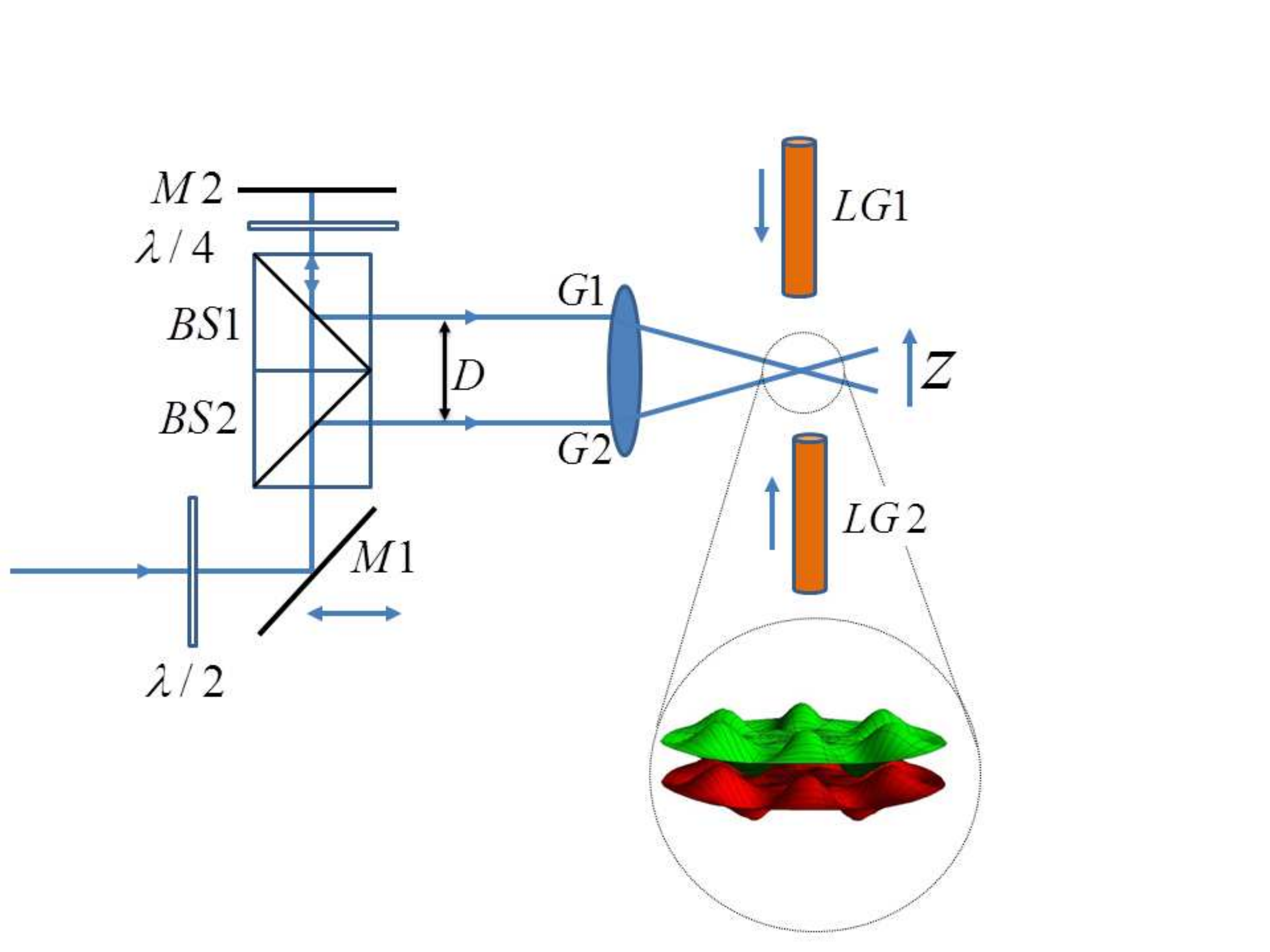}}
\end{center}
\caption{(Color
online). Proposed setup for the ring-ring coupling. Two gaussian laser beams of wavelength $\lambda$ and distance $D$, pass through a lens and
 interfere in the focal plane ($f$ is the focal length). The distance $D$ can be easily controlled by moving the mirrors. The distance between the fringes is a function of $1/D$ \cite{Li}. The resulting Gaussian laser beam with wave vector ${k_{G}={2 \pi D}/{(\lambda f)}} $, then, interferes with two counter propagating Laguerre-Gauss laser beams of amplitude $E_0$. The inset shows the ring lattice potentials separated by $d=\lambda {f}/{D}$. Here $l=6$ and $p=0$. }
 \label{double-ring potential}
\end{figure}

In  this section, we describe the experimental setup  for realizing two  ring-shaped optical lattices with a tunable interaction between of them.  To achieve the task, we use  Laguerre-Gauss (LG) modes to produce closed optical lattices\cite{hologram,LG gen}. The electric field, with frequency $\omega$, wave vector $\textit{k}$ and amplitude $E_{0}$, which is propagating along the $\textit{z}$ axis, can be written in cylindrical coordinates $(r,\phi,z)$ as
$E(r,\phi,z)=E_{0}f_{pl}(r)e^{i l \phi} e^{i(\omega t-kz )}$, $f_{pl}(r)=(-1)^{p}\sqrt{\frac{2p!}{\pi(p+|l|)!}}\varepsilon^{l}L_{p}^{|l|}(\varepsilon^{2})e^{-\varepsilon^{2}}$, $\varepsilon=\sqrt{2}r/r_{0}$, where
$r_{0}$ is the waist of the beam and $L_{p}^{|l|}$ are associated Laguerre polynomials $L_{p}^{|l|}(z)=(-1)^{m}d^{m}/dx^{m}[L_{n+m}(z)]$, with $L_{n+m}(z)]$ being the Laguerre polynomials. The numbers $p$ and $l$ label the radial and azimuthal quantum-coordinates. In this paper we will consider  the case of $p=0$.
We realize an adjustable distance between the two rings by
 changing the standing wave periodicity in a controllable way \cite{Li}: Two beams which are passing through lens
 interfere with each other at the focal plane and create
 interference pattern; the
 periodicity of the obtained lattice is  inversely proportional to the distance
 between the two beams. 
 The  set-up  is depicted in  the Fig.~\ref{double-ring potential}. The potential which is obtained at the focal plane of lenses has the following form:

\begin{eqnarray}\nonumber
V_{latt}=4E_{0}^{2}(f_{l}^{2} \cos{(k_{LG}z)}^2+\cos{(k_{G}z)}^2+\\
2f_{l} \cos{(k_{LG} z)}\cos{(k_{G} z)} \cos{(\phi l)}),
\label{potential}
\end{eqnarray}
where $k_{LG}$ is the  wave-vector of the Laguerre-Gauss beam.  $k_{G}=\frac{2 \pi D}{\lambda f} $ is the  effective wave vector for the Gaussian beams, where  $D$, $\lambda$ and $f$  are  respectively the distance between the two beams passing through the lens, the wavelength of the Gaussian beams and  the focal length of the lens.  Using this equation,  we can conclude that a stack of closed rings is initially obtained, with $N=l$ lattice sites. The depth of the  wells along each ring scales as $\sqrt{1/l!}$. The distance between rings can be  controlled by changing the distance between the two Gaussian beams. This  can be  realized  by moving the mirror $M1$. It can be shown that when $k_{G}=k_{LG}$,  Eq.(\ref{potential}) will give the potential obtained in \cite{Luigi}. So, this potential can be regarded as a generalization of previously obtained potential for the ring-shaped optical lattices.
The tunneling matrix element between two rings in the limit $V_0\gg E_r$, where $V_0=4E_{0}^{2}$ end $E_r=\frac{\hbar^2 k^2}{2m}$ is the recoil energy, is given by
\be
g=4\sqrt{\frac{\hbar}{\sqrt{2m}}}\frac{V_0^{3/4}}{\sqrt{d}}e^{-\frac{\sqrt{2m V_0}}{\pi\hbar}d}
\ee
here $d=\lambda f/D$ is the lattice spacing along z-direction.
The physical parameters of the set-up are summarized as follows.
With a laser intensity of $I = 5W/cm^2$ and the detuning $\Delta = -10^6MHz$ the potential wells are separated
by a barrier of $\sim5\mu K$ much larger than the chemical potential of a standard condensate
(whose temperature can reach few nK).With these parameters the scattering rate is $\sim1$ photon/sec.
It is feasible to have a ring lattice with a radius  $r_0\sim 5 \mu m$ and  $N= 20$ lattice sites. With the laser wavelength $\lambda=830 nm$ and lens with $f=40mm$  the separation between two rings can be adjustable with the setup shown in Fig.~\ref{double-ring potential} in a range of $1.7$ to  $6$ $\mu m$, by changing $D$ from 19.6 to 5.5$mm$, whereby the lattice well spacing within each ring is  $\sim 1.57$ $\mu m$.

Assuming that the particles  occupy the lowest Bloch level only (low temperature) ,  both the intra-ring and inter-rings  tunneling amplitudes will have a negligible  dependence on the ring's radial coordinates.

A single-species bosonic condensate  is envisaged to be loaded in the setup described above.  Our system is thus governed by a Bose-Hubbard ladder Hamiltonian:
\begin{equation}
 H_{BH}=H_{a}+H_{b}+H_{int} - \sum_{\alpha=a,b}\sum_{i=0}^{N-1}\mu_{\alpha}\hat{n}_{i}^{\alpha}
 \label{model}
 \end{equation}
 with
\begin{eqnarray}
&&H_{a}=-t\sum_{i=0}^{N-1}(e^{i\Phi_{a}/N}a_{i}^{\dag}a_{i+1}+h.c.)+\frac{U}{2}\sum_{i=1}^{N}\hat{n}_{i}^{a}(\hat{n}_{i}^{a}-1) \nonumber  \\
%
&&H_{b}=-t\sum_{i=0}^{N-1}(e^{i\Phi_{b}/N}b_{i}^{+}b_{i+1}+h.c.)+\frac{U}{2}\sum_{i=1}^{N}\hat{n}_{i}^{b}(\hat{n}_{i}^{b}-1) \nonumber \\
 %
 &&H_{int}=-g\sum_{i=0}^{N-1}(a_{i}^{\dag}b_{i}+b_{i}^{\dag}a_{i})
 \label{model}
 \end{eqnarray}
where $H_{a,b}$ are the Hamiltonians of the condensates in the rings $a$ and $b$ and the $H_{int}$ describes the interaction between rings. Operators $\hat{n}_{i}^{a}=a_{i}^{\dag}a_{i},
\hat{n}_{i}^{b}= b_{i}^{\dag}b_{i}$ are the particle number operators for the lattice site $i$. Operators $a_{i}$ and $b_{i}$ obey to the standard bosonic commutation relations . The parameter $t=\int \textit{w}^{a,b}(\textbf{x}-\textbf{x}_{i}^{a,b})(-\frac{\hbar^2}{2m}\nabla^2+V_{latt})\textit{w}^{a,b}(\textbf{x}-\textbf{x}_{i}^{a,b})$ is the tunneling rate within lattice neighboring sites(in the rings a and b), and $g=\int \textit{w}^{a}(\textbf{x}-\textbf{x}_{i}^{a})(-\frac{\hbar^2}{2m}\nabla^2+V_{latt})\textit{w}^{b}(\textbf{x}-\textbf{x}_{i}^{b}) d^3\textbf{x}$ is the tunneling rate between the rings, where $\textit{w}^{a,b}(\textbf{x})$ and $\textbf{x}_{i}^{a,b}$ are the single particle Wannier functions and site index for the rings a and b respectively and $\textit{w}^{a,b}(\textbf{x}-\textbf{x}_{i}^{a,b})=w(x-x_i)w(y-y_i)w(z-z_i \pm d/2)$(where $+$ sign holds for the ring \emph{a} and $-$ sign for the ring \emph{b}), with $d$ being distance between the rings. Here we assume that Wannier functions for the two rings are the same. The repulsion between two atoms on single lattice site is quantified by the on-site
interaction matrix element $U=\frac{4\pi a_{s} \hbar^{2}}{m}\int |\textit{w}^{a,b}(\textbf{x})|^4 d^3\textbf{x}$, where $a_{s}$ is the $s$-wave scattering length of an atom.
Finally, the phases $\Phi_a$ and $\Phi_b$ are the phase twists responsible for the currents flowing along the rings.
%
They can be expressed through vector potential of the so-called synthetic gauge fields in the following way: $\Phi_{a}/N=\int_{x_{i}}^{x_{i+1}}\emph{\textbf{A(z)}}d\emph{\textbf{z}}$, $\Phi_{b}/N=\int_{x_{i}}^{x_{i+1}}\emph{\textbf{B(z)}}d\emph{\textbf{z}}$, where $\emph{\textbf{A(z)}}$ and $\emph{\textbf{B(z)}}$ are generated vector potentials in the rings $a$ and $b$, respectively (see appendix\ref{peierls}).

\section{Effective qubit  dynamics}\label{sec:qubit_dynamics}
In this section, we demonstrate that  the effective phase dynamics of the system indeed defines a qubit.
To this end, we elaborate on the imaginary-time path integral of the partition function of the model Eq.(\ref{model}) in the limit of large fluctuations of the number of bosons at each site.  We first perform a local gauge transformation
$a_l\rightarrow  a_l e^{i l \Phi_a}$, $b_l\rightarrow  b_l e^{i l \Phi_b}$, eliminating the contribution of the magnetic field everywhere except at a given site  of the ring (twisted boundary conditions\cite{twisted boundary}). In the regime under scrutiny, the partition function of the model Eq.(\ref{model}) is \cite{Feynman1,Fazio}
\begin{equation}
Z=Tr \left ( e^{-\beta H_{BH}} \right )  \propto \int D[\{ \phi_i \}]  e^{-S[\{ \phi_i \}]}
\end{equation}
where the effective action is
\begin{widetext}
\begin{eqnarray}
&&S[\{\phi_i\}]=S_0[\{\phi_i\}]+S_{int}[\{\phi_i\}]\\
&&S_0[\{\phi_i\}]=\int_0^\beta d\tau \sum_{i=0 \atop \alpha=\{ a,b\} }^{N-2}  \left [ {\frac{1}{U}} (\dot{\phi}_{i,\alpha})^2-E_J \cos\left ( \phi_{i+1,\alpha}-\phi_{i,\alpha}\right )\right ]   + \left [ {\frac{1}{U}} (\dot{\phi}_{N-1,\alpha})^2-E_J \cos\left ( \phi_{0,\alpha}-\phi_{N-1,\alpha}-\Phi_\alpha\right )\right ]  \\
&&S_{int}[\{\phi_i\}]= - E_J'\int_0^\beta  d\tau  \sum_{i=0 }^{N-1}    \cos \left ( \phi_{i,a}-\phi_{i,b}-{\frac{\Phi_a-\Phi_b}{N}} i \right )
\end{eqnarray}
\end{widetext}
with $E_J=t \langle n \rangle $ and $E_J'=g \langle n \rangle $.

Because of the gauge transformations, the phase slip is produced only at the boundary. We define $\theta_\alpha\doteq \phi_{N-1,\alpha}-\phi_{0,\alpha}$. The goal now is to integrate out the phase variables in the bulk. Assuming  that the two rings are weakly coupled and that $U/E_J<<1$,  the bulk variables are not involved in the inter-ring tunneling term because we can take $\phi_{i,a}\approx \phi_{i,b}$ everywhere except at the boundary:
\begin{eqnarray}
\sum_{i=0}^{N-1} && \cos  \left ( \phi_{i,a}-\phi_{i,b} -{\frac{\Phi_a-\Phi_b}{N}} i \right )=\sum_{i=0}^{N-2} \cos \left ( {\frac{\Phi_a-\Phi_b}{N}} i \right ) \nonumber \\  &&+ \cos \left ( \theta_a-\theta_b -{\frac{\Phi_a-\Phi_b}{N}} (N-1) \right )
\end{eqnarray}
where, without loss of generality,   we can assume  $\phi_{0,a}\equiv \phi_{0,b}$. Therefore the non-trivial path integration corresponds to $S_0[\{\phi_i\}]$ only.
To achieve the task we observe that in the phase-slips-free-sites the phase differences are small, so the harmonic approximation can be applied:
\begin{eqnarray}
\sum_{i=0}^{N-1} \cos\left ( \phi_{i+1,\alpha}-\phi_{i,\alpha}\right ) \mapsto  && \cos( \theta_{\alpha}-\Phi_\alpha) - \\  &&\sum_{i=0}^{N-2}   {{\left ( \phi_{i+1,\alpha}-\phi_{i,\alpha}\right )^2}\over{2}}  \nonumber \;.
\end{eqnarray}
In order to facilitate the integration in the bulk phases, we express the single $\phi_{0,\alpha}$ and $\phi_{N-1,\alpha}$ as: $\phi_{0,\alpha}=\tilde{\phi}_{0,\alpha}+\theta_\alpha/2$, $\phi_{N-1,\alpha}=\tilde{\phi}_{0,\alpha}-\theta_\alpha/2$. We observe that the sum of the quadratic terms above involves $N-1$ fields with periodic boundary conditions: $\{\tilde{\phi}_{0,\alpha}, \phi_{1,\alpha},\dots,  \phi_{N-2,\alpha} \} \equiv \{ \psi_{0,\alpha},\psi_{1,\alpha},\dots, \psi_{N-2,\alpha}\}$, $\psi_{N-1,\alpha}=\psi_{0,\alpha}$. Therefore,
\begin{eqnarray}
\sum_{i=0}^{N-2}  \left ( \phi_{i+1,\alpha}-\phi_{i,\alpha}\right )^2=\sum_{i=0}^{N-2}  \left ( \psi_{i+1,\alpha}-\psi_{i,\alpha}\right )^2 + \nonumber \\
{{1}\over{2}}\theta_\alpha^2 +\theta_\alpha \left (\psi_{N-2,\alpha}-\psi_{1,\alpha} \right ) \;.
\end{eqnarray}

The effective action, $S_0[\{\phi_i\}]$, can be split into two terms
$
S_0[\{\phi_i\}]=S_{01}[\theta_\alpha] +S_{02}[\{\psi_{i\alpha}\}]
$
with
\begin{widetext}
\begin{eqnarray}
&&S_{01}[\theta_\alpha]=\int_0^\beta d\tau \left [ {\frac{1}{U}} (\dot{\theta}_{\alpha})^2+ {{E_J}\over{2}} \theta_\alpha^2-E_J \cos\left (\theta_\alpha -\Phi_\alpha \right )\right ]  \label{S_theta}\\
&& S_{02}[\{\psi_{i,\alpha}\}, \theta_\alpha]=\int_0^\beta d\tau \left \{{\frac{1}{U}} (\dot{\psi}_{0,\alpha})^2+  \sum_{i=0}^{N-2} \left [ \frac{1}{U} (\dot{\psi}_{i,\alpha})^2 +{{E_J}\over{2}}   \left ( \psi_{i+1,\alpha}-\psi_{i,\alpha}\right )^2\right ] + E_J  \theta_\alpha \left (\psi_{N-2,\alpha}-\psi_{1,\alpha} \right )  \right \}
\label{S_coupling}
\end{eqnarray}
\end{widetext}
The integration of the fields $\psi_{i,\alpha}$ proceeds according to the standard methods (see \cite{Hekking}). The fields that need to be integrated out are expanded in Fourier series ($N$ is assumed to be even):
$\psi_{l,\alpha}=\frac{1}{\sqrt{N-1}}\left [\psi_{0,\alpha}+(-)^l \psi_{N/2,\alpha}+\sum_{k=1}^{(N-2)/2} \left (\psi_{k,\alpha} e^{{{2\pi i k l}\over{N-1}}}+c.c. \right )\right ]$,  with $\psi_{k,\alpha}=a_{k,\alpha}+ib_{k,\alpha}$. The coupling term in Eq. (\ref{S_coupling}) involves only the imaginary part of $\psi_{k,\alpha}$: $ \psi_{N-2,\alpha}-\psi_{1,\alpha}=\sum_k b_{k,\alpha} \zeta_k$, being  $\displaystyle{\zeta_k={{4}\over{\sqrt{N-1}}} \sin \left ( {{2 \pi k}\over{N-1}}\right ) } $. Therefore:
\begin{widetext}
\begin{equation}
S_{02}[\{\psi_{i,\alpha}\}, \theta_\alpha] =\int_0^\beta d\tau {{1}\over{U}}\sum_k \left[(\dot{a}_{k,\alpha})^2 +\omega_k^2 a_{k,\alpha}^2 \right]+
\int_0^\beta d\tau {{1}\over{U}}\sum_k \left[(\dot{b}_{k,\alpha})^2 +\omega_k^2 b_{k,\alpha}^2+ E_J U \zeta_k\theta_\alpha   b_{k,\alpha}\right]
\end{equation}
\end{widetext}
where $ \omega_k= \sqrt{2 E_J U \left [ 1-\cos{ \left ( {{2\pi k}\over{N-1}} \right )} \right ] } $. The integral in $\{a_{k,\alpha} \} $ leads to a Gaussian path integral; it does not contain the interaction with $\theta_\alpha$, and therefore brings a prefactor multiplying the effective action, that does not affect the dynamics. The integral in $\{b_{k,\alpha} \} $ involves  the interaction and therefore leads to a non local kernel in the imaginary time: $\int d\tau d\tau' \theta_\alpha(\tau) G(\tau-\tau')\theta_\alpha(\tau')$. The explicit form of $G(\tau-\tau') $ is obtained by expanding  $\{b_{k,\alpha} \} $ and $\theta_\alpha$ in Matsubara frequencies $\omega_l$. The corresponding Gaussian integral yields  to the
\begin{equation}
\int D[b_{k,\alpha}] e^{-\int_0^\beta d \tau S_{02}}\propto \exp{\left (-\beta U E_J^2 \sum_{l=0}^{\infty} \tilde{Y}(\omega_l) |\theta_l|^2  \right )}
\end{equation}
with $\tilde{Y}(\omega_l)=\sum_{k=1}^{(N-2)/2} {{\zeta_k^2}\over{\omega_k^2+\omega_l^2}}$. The $\tau=\tau'$ term is extracted by summing and subtracting $\tilde{Y}(\omega_l=0)$; this compensates the second term in Eq.(\ref{S_theta}).

\begin{figure}
{\includegraphics*[scale = 0.6]{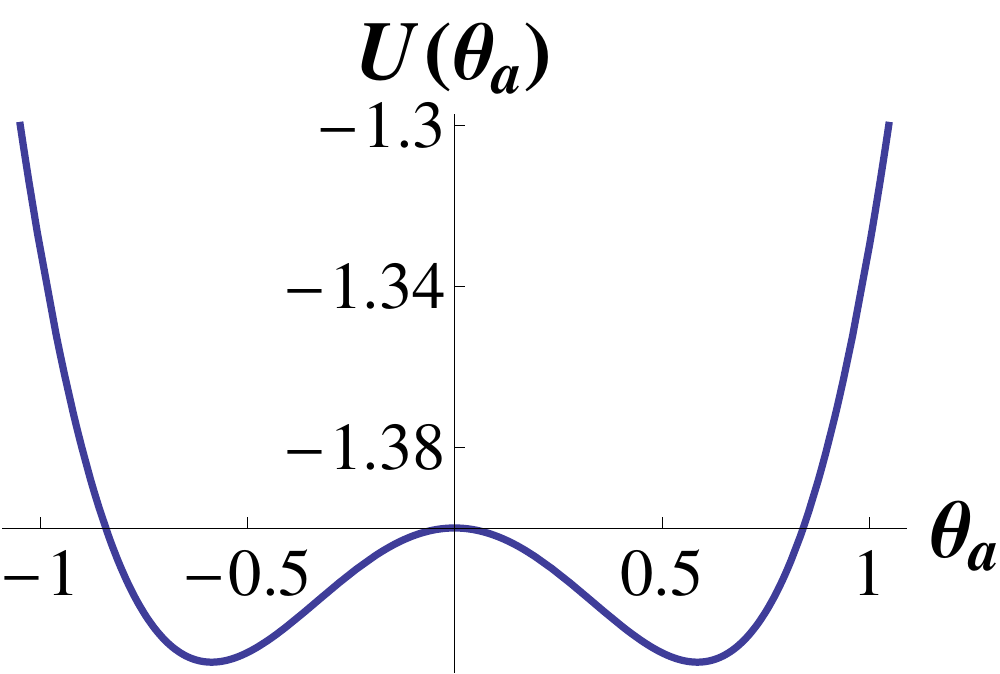}}
{\includegraphics*[scale = 0.6]{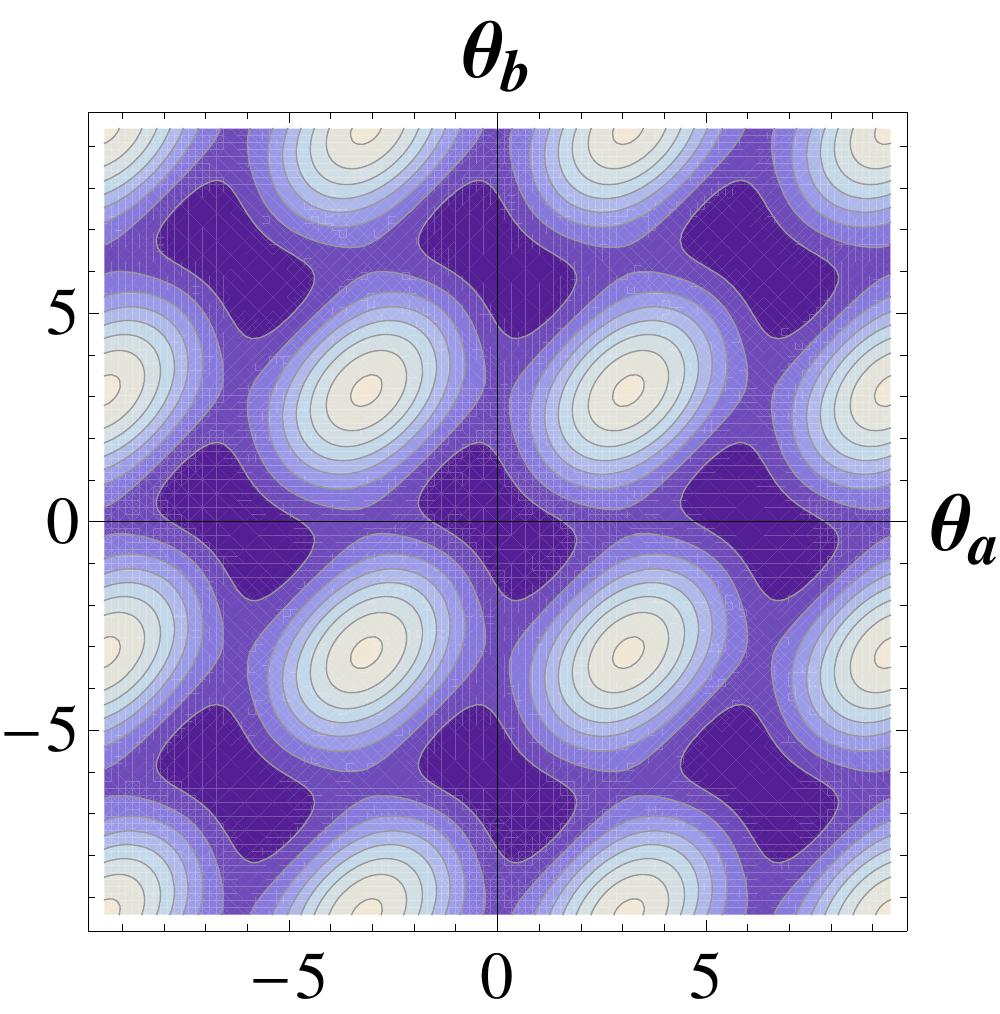}}
\caption{(Color
online). Top: The effective potential landscape. Bottom:  The double well for $\theta_a=-\theta_b$. The parameters are $E_J'/E_J=0.8$ and $ \Phi_a-\Phi_b=\pi$.}
\label{contourplot}
\end{figure}

The effective action finally reads as
\begin{widetext}
\begin{equation}
S_{eff}=\int_0^\beta d \tau \left [ {\frac {1}{2 U}}  \sum_{\alpha=a,b} \dot{\theta}_\alpha^2 + U(\theta_a,\theta_b) \right ]
- \frac{E_J}{2 U(N-1)} \sum_{\alpha=a,b}   \int_0^\beta d \tau d \tau' \theta_\alpha(\tau)G_\alpha(\tau-\tau') \theta_\alpha(\tau')
\end{equation}
\end{widetext}
where
\begin{eqnarray}
U(\theta_a,\theta_b)\doteq && \sum_{\alpha=a,b}  \left [ \frac{E_J}{2(N-1)} (\theta_\alpha-\Phi_\alpha)^2 -E_J \cos(\theta_\alpha) \right ] \nonumber  \\
&& - {E_J'}  \cos[\theta_a-\theta_b-{\frac{N-2}{N}}(\Phi_a-\Phi_b)]  \; .
\end{eqnarray}
We observe that for large $N$, the potential  $U(\theta_a,\theta_b)$ provides  the  effective phase dynamics of Josephson junctions flux qubits realized by Mooij {\it et al.}  (large $N$ corresponds  to large geometrical inductance of flux qubit devices) \cite{Mooij}. In that article, the landscape was thoroughly analyzed.  The qubit is realized by superposing  the two states $|\theta_1 \rangle $ and $|\theta_2 \rangle $  corresponding to the minima of  $U(\theta_a,\theta_b)$. The degeneracy point is achieved at  $\Phi_b-\Phi_a=\pi$(See Fig.\ref{contourplot}). We comment that the ratio  ${E_J'}/E_J $ controls the relative size of the energy barriers between  minima intra- and minima inter-'unit cells' of the $(\theta_a,\theta_b)$ phase space, and therefore is important for designing the qubit.  In our system $E_{J}'/E_J $ can be fine tuned with the scheme shown in Fig.\ref{double-ring potential}.
The kernel in the non-local term  is given by $G_a(\tau)=G_b(\tau)=G(\tau)$, with:
\begin{equation}
G(\tau)=\sum_{l=0}^\infty  \sum_{k=1}^{{{N-2}\over{2}}} {\frac{\omega_l^2 \left (1+\cos[{{2 \pi k}\over{N-1}}]\right )}{2 E_JU(1-\cos[{{2 \pi k}\over{N-1}}])+\omega_l^2}} e^{i\omega_l \tau}\; .
\end{equation}
The external bath vanishes in the thermodynamic limit and the effective action reduces to the Caldeira-Leggett one \cite{Hekking,Legget}.

\section{Real time dynamics: Two  coupled Gross-Pitaevskii equations}\label{sec:real_dynamics}

In this section we study the dynamics of the  number and phase imbalance  of  two bose-condensates  confined  in the ring shaped potential.  For this goal, we assume that the system is described by a Bose-Hubbard ladder Eqs.(\ref{model}), is in a superfluid  regime, with   negligible quantum fluctuations.
The order parameters can be defined as the expectation values of boson operators in the Heisenberg picture:
\begin{eqnarray}
 \varphi_{a,i}(s)=\langle a_{i}(s)\rangle,\varphi_{b,i}(s)=\langle b_{i}(s)\rangle
 \label{app0}
\end{eqnarray}
implying that the  Heisenberg equations for  the operators $a_{i}$ and $b_{i}$ are simplified into the  Gross-Pitaevskii equations for the corresponding expectation values:
\begin{eqnarray}\nonumber
i\hbar\frac{\partial \varphi_{a,i} }{\partial s}=-t(e^{i\Phi_{a}/N}\varphi_{a,i+1}+e^{-i\Phi_{a}/N}\varphi_{a,i-1}) \\
+U|\varphi_{a,i}|^{2}\varphi_{a,i}-\mu_{a}\varphi_{a,i}-g\varphi_{b,i}
\label{eq1}
\end{eqnarray}
\begin{eqnarray}\nonumber
i\hbar\frac{\partial \varphi_{b,i} }{\partial s}=-t(e^{i\Phi_{b}/N}\varphi_{b,i+1}+e^{-i\Phi_{b}/N}\varphi_{b,i-1}) \\
+U|\varphi_{b,i}|^{2}\varphi_{b,i}-\mu_{b}\varphi_{b,i}-g\varphi_{a,i}
\label{eq2}
\end{eqnarray}
We assume that
$\varphi_{a,i+1}-\varphi_{a,i}=\frac{\varphi_{a}(s)}{\sqrt{N}}$  and  $\varphi_{b,i+1}-\varphi_{b,i}=\frac{\varphi_{b}(s)}{\sqrt{N}}$
for all $i,j=0,..,N$,
where $N$ is a total number of ring-lattice sites.
From Eqs.(\ref{eq1}) and (\ref{eq2}) we obtain
\begin{eqnarray}\nonumber
i\hbar\frac{\partial \varphi_{a} }{\partial s}=-2t\cos{(\Phi_{a}/N)}\varphi_{a}
+\frac{U}{N}|\varphi_{a}|^{2}\varphi_{a} \\
-\mu_{a}\varphi_{a}-g\varphi_{b}
\label{eq3}
\end{eqnarray}
\begin{eqnarray}\nonumber
i\hbar\frac{\partial \varphi_{b} }{\partial s}=-2t\cos{(\Phi_{b}/N)}\varphi_{b}
+\frac{U}{N}|\varphi_{b}|^{2}\varphi_{b} \\
-\mu_{b}\varphi_{b}-g\varphi_{a}
\label{eq4}
\end{eqnarray}
Employing the standard phase-number representation:  $\varphi_{a,b}=\sqrt{N_{a,b}} e^{i\theta_{a,b}}$,  two pairs of equations are obtained for imaginary and real parts:
\begin{eqnarray}\nonumber
\hbar\frac{\partial N_{a} }{\partial s}=-2g\sqrt{N_{a}N_{b}}\sin{(\theta_{b}-\theta_{a})}
\end{eqnarray}
\begin{eqnarray}
\hbar\frac{\partial N_{b} }{\partial s}=2g\sqrt{N_{a}N_{b}}\sin{(\theta_{b}-\theta_{a})}
\label{eq5}
\end{eqnarray}
\begin{eqnarray}\label{eq6}
\hbar\frac{\partial \theta_{a} }{\partial s}=-2t\cos{\Phi_{a}/N}-\frac{U N_{a}}{N}+\mu_{a}+g\sqrt\frac{N_{b}}{N_{a}}\cos{(\theta_{b}-\theta_{a})} \nonumber \\
\hbar\frac{\partial \theta_{b} }{\partial s}=-2t\cos{\Phi_{b}/N}-\frac{U N_{b}}{N}+\mu_{b}+g\sqrt\frac{N_{a}}{N_{b}}\cos{(\theta_{b}-\theta_{a})} \nonumber \\
\end{eqnarray}
From Eqs.(\ref{eq5}) it results that $\frac{\partial N_{a} }{\partial s}+\frac{\partial N_{b} }{\partial s}=0$, reflecting the conservation of the total bosonic number $N_T=N_a+N_b$.
From equations (\ref{eq5}) and (\ref{eq6}) we get
\begin{equation}
\frac{\partial Z }{\partial \tilde {s}}=-\sqrt{1-Z^{2}}\sin{\Theta}
\label{dynamical_equation1}
\end{equation}
\begin{equation}
\frac{\partial \Theta }{\partial \tilde {s}
}=\Delta+\lambda \rho Z+\frac{Z}{\sqrt {1-Z^{2}}}\cos{\Theta}
\label{dynamical_equation2}
\end{equation}
where we introduced  new variables:the dimensionless time $ 2gs/\hbar\rightarrow \tilde{s}$,the population imbalance $Z(\tilde{s})=(N_{b}-N_{a})/(N_{a}+N_{b})$ and the phase difference between the two condensates $\Theta(\tilde{s})=\theta_{a}-\theta_{b}$.
It is convenient to characterize system with a new set of parameters:external driving force $\Delta=(2t(\cos{\Phi_{a}/N}-\cos{\Phi_{b}/N})+\mu_{b}-\mu_{a})/2g$, the effective scattering wavelength $\lambda =U/2g$ and the total bosonic density $\rho=N_{T}/N$.
The exact  solutions of  Eqs.(\ref{dynamical_equation1}) and (\ref{dynamical_equation2})  in terms of elliptic functions\cite{Smerzi}  can be adapted to our case and it is detailed in the Appendix \ref{elliptic}. As it has been noticed in \cite{Smerzi}  equations can be derived as Hamilton equations with
\begin{equation}
H(Z(\tilde {s}),\Theta(\tilde {s}))=\frac{\lambda \rho Z^2}{2}+\Delta Z-\sqrt{1-Z^2}cos\Theta,
\label{eff ham}
\end{equation}
by considering $Z$ and $\Theta$ as conjugate variables.
Since  the energy of the   system is conserved,   $H(Z(\tilde {s}),\Theta(\tilde {s}))=H(Z(0),\Theta(0))=H_{0}$.

The dynamics can be visualized with the help of the mechanical system provided by a rotator of length of $\sqrt{1-Z^2}$ driven by the external force $\Delta$. In this picture, $Z$ is considered to be an angular momentum of the rotator and $\lambda \rho$ its moment of inertia. For $\Delta=0$,  the dynamics  of the rotator depends on the value of initial angular momentum $Z(0)$. For small  initial kinetic energy $\frac{\lambda \rho Z(0)^2}{2}$, the  rotator makes small oscillations and its trajectory is an opened curve, thus $\left \langle Z \right \rangle=0$. For the critical value $Z_c$ the rotator reaches the equilibrium  vertical position corresponding to $\phi=\pi$. For $Z_0>Z_c$ the rotator performs revolutions   around his fixed point with $\Delta \theta =2\pi$ and $\left \langle Z \right \rangle=0$. In the case of non vanishing $\Delta$,  the dynamics of the rotator depends on $Z(0)$ and external force $\Delta$. Because of the $\Delta$, the  system oscillates  around a shifted equilibrium value (which leads to an  asymmetry into the system) and $\left \langle Z \right \rangle\neq0$ for all the cases.

Below, we discuss the  different regimes  for population imbalance, emerging from  Eqs.(\ref{dynamical_equation1}) and (\ref{dynamical_equation2}) depending on the values of parameters $\lambda \rho$ and $Z_0$ (see the  Appendix \ref{elliptic} for technical details). For each physical regime we further discuss the solution for both  cases of vanishing and non vanishing $\Delta$.

\subsection{\protect\normalsize Population imbalance and oscillation frequencies  in the limit $\lambda \rho=0$ }
{ I-A} $\Delta=0.$--
For noninteracting atoms, the  solution of Eqs.(\ref{dynamical_equation1}) and (\ref{dynamical_equation2}) is
\begin{equation}
Z(\tilde {s})=\sqrt{1-H_{0}^2}\sin{(\tilde {s}+\tilde {s}_{0})}
\label{sol0}
\end {equation}
where $\tilde {s}_{0}=\arcsin{\frac{Z_{0}}{\sqrt{1-H_{0}^2}}}$ and $H_0=-\sqrt{1-Z_{0}^2}\cos{\Theta_{0}}$ is an initial energy of the system.
Eq.(\ref{sol0}) describes sinusoidal
Rabi oscillations between the two traps with frequency $\omega_{0} = 2g$. These oscillations are
equivalent to single atom dynamics, rather than a Josephson-effect arising from the interacting
superfluid condensate.

{ II-A} $\Delta\neq0.$ --
Depending on the  value of the determinant $D=1-H_0^{2}/(\Delta^2+1)$ of the equation $f(Z)=-(\Delta^2+1)Z^2+2H_0\Delta Z+1-H_0^2$,  the population imbalance is either oscillating around a non-zero average, reflecting the  MQST phenomenon,  or staying constant in time.
A numerical analysis shows that $D\geq0$.
Therefore, there are two different sub-cases.
When $D=0$ (which can be satisfied only if $\sin{\Theta_0}=0$), the population imbalance stays constant and takes on the value
\begin{equation}
Z=Z_0=-\frac{\Delta}{\sqrt{1+\Delta^2}}=const
\label{sol D=0}
\end{equation}
For this value of  the initial population imbalance, $\Theta=const$.
In the case when $D>0$,
the subsequent  expression for $Z(t)$ is obtained:
\begin{equation}
Z(\tilde {s})=B-\frac{C}{A}\sin{[a(\tilde {s}-\tilde {s}_{0})]},
\label{sol0.1}
\end{equation}
where $\tilde {s}_{0}=\frac{1}{A}\arcsin{[\frac{A}{C}(Z_{0}-B)]},A=\sqrt{1+\Delta^2},B=\frac{\Delta H_0}{A^2}$ and $C=\sqrt{1-\frac{H_0^2}{A^2}}$.
As it is seen from Eq.(\ref{sol0.1}), the  system is oscillating about a non-zero average value $B$ with frequency
\begin{equation}
\omega=\omega_{0}\sqrt{1+\Delta^2}
\end{equation}

\begin{figure}
\begin{center}
{\includegraphics*[scale = 0.6]{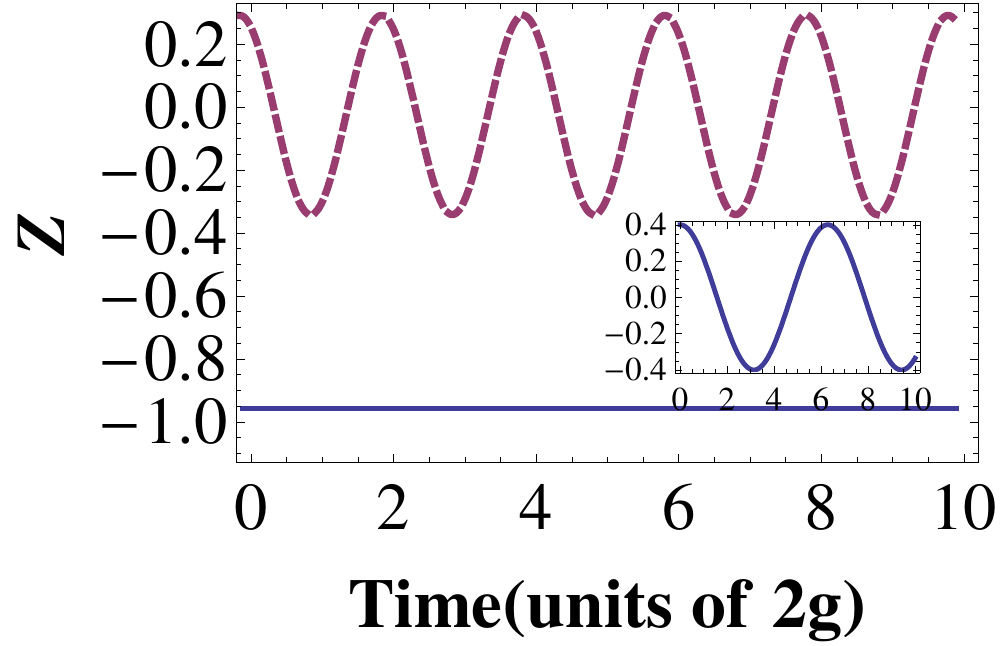}}
\end{center}
\caption{(Color
online). The population imbalance  in  two coupled rings for the case  $\lambda \rho=0$.
Solid and dashed lines correspond to the cases $D=0$ and $D>0$ accordingly in the regime of non-zero $\Delta$.
An inset shows behavior of $Z(\tilde{s})$ for vanishing $\Delta$.
Here
$\lambda\rho = 0$, $\Delta = 2$, $\Theta_{0}=0$ implying that $\omega \simeq 3.16\omega_{0}$.}
\label{graph liambda=0}
\end{figure}

All the regimes discussed for this case are displayed in Fig.\ref{graph liambda=0}.

\subsection{\protect\normalsize Population imbalance and oscillation frequencies   in the limit $\lambda \rho\ll1$ }

\begin{figure}
\begin{center}
{\includegraphics*[scale = 0.6]{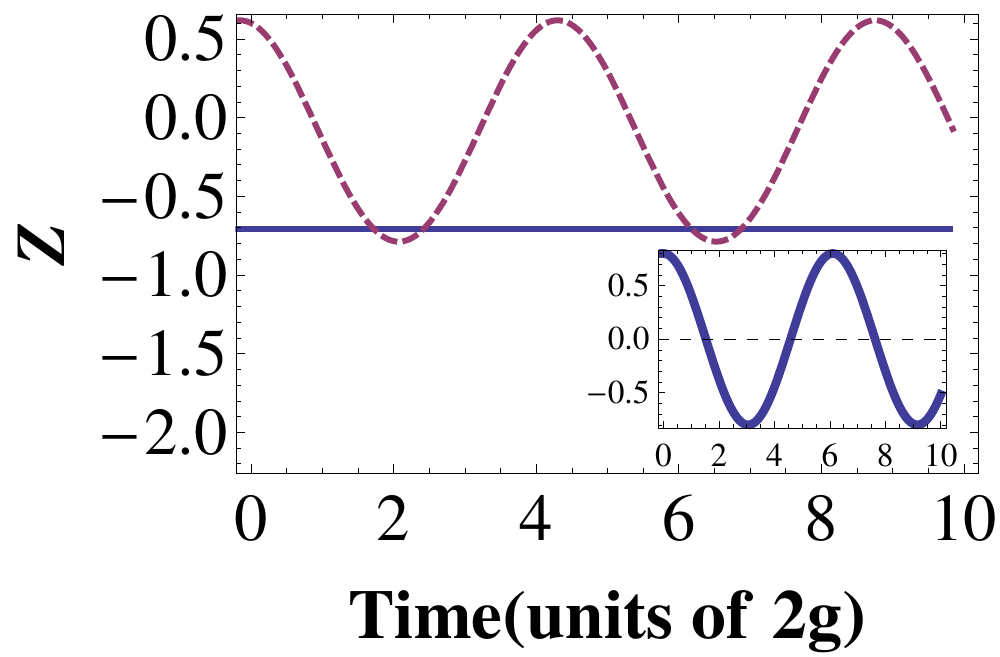}}
\end{center}
\caption{(Color
online). The population imbalance  in  two coupled rings for the case  $\lambda\rho\ll1$.
Solid and dashed lines correspond to the cases $\delta=0$ and $\delta<0$ accordingly in the regime of non-zero $\Delta$.
An inset shows behavior of $Z(\tilde{s})$ for the vanishing $\Delta$. Here
$\lambda\rho = 0.1$ , $\Theta_{0}=0$ and $\Delta = 1$.}
\label{graph liambda=0.1}
\end{figure}

{I-B} $\Delta=0.$--
The qualitative behavior of the dynamics for this subcase depends  on the elliptic modulus  $k$ which is given by Eq.(\ref{el mod}).
 For $\lambda \rho\ll 1$
 \begin{equation}
k=Z(0)\lambda \rho(1-\frac{\lambda \rho}{2}\sqrt{1-Z(0)^2})
\end{equation}
So $ k\sim 0$ and  therefore $Z(t)$ displays only one regime given by:
 \begin{eqnarray}
Z(\tilde {s}) &
\simeq & Z(0)(\cos{\omega (\tilde {s}-\tilde {s}_{0})} \\ \nonumber
&+&   \frac{k}{4}(\omega (\tilde {s}-\tilde {s}_{0})-\sin{2 \omega (\tilde {s}-\tilde {s}_{0})})
\sin{ \omega (\tilde {s}-\tilde {s}_{0})})
\end{eqnarray}
 where $\omega\simeq2g(1+\frac{\lambda}{2} \rho\sqrt{1-Z(0)^2})$ and $\tilde {s}_{0}$ is  fixing initial condition.
Therefore, in this regime the population imbalance is characterized by  almost sinusoidal oscillations about zero average-- see the inset of Fig.\ref{graph liambda=0.1}.

{ II-B} $\Delta\neq0.$--
In this case behavior of $Z(t)$ is governed by determinant $\delta$ of the  cubic equation Eq.(\ref{discrim}).
There are two different regimes depending on the initial value of the population imbalance which are given by the value of $\delta$.
All the regimes can be discussed by expressing the Weierstrass function in Eq.(\ref{weier sol}) using Jacobian elliptic functions.
In the limit of $\delta=0$, the population imbalance takes form
\begin{equation}
Z(\tilde {s})=Z(0)+\frac{f'[Z(0)]/4}{-c+3c[\sin
{(-\sqrt{3c}\frac{\lambda \rho}{2} \tilde {s}})]^{-2}-f''[Z(0)]/24}
\end{equation}
For the parameters discussed in the article, $ f'[Z(0)]\sim10^{-14}$; so the population imbalance is staying constant due to the same reason as for the subcase $D=0$ of the previous section.
In the limit of $\delta<0$, the population imbalance takes form
\begin{equation}
Z(\tilde {s})=Z(0)+\frac{f'[Z(0)]/4}{e_{2}+H_{2}\frac{1+\cos{(\lambda \rho  \sqrt{H_{2}\tilde {s}}})}{1-\cos{(\lambda \rho  \sqrt{H_{2}\tilde {s}}})}-f''[Z(0)]/24}
\label{saturation}
\end{equation}
where $e2,H_{2}$ are defined in the Appendix B. Eq.(\ref{saturation}) is correct when $1/2-3e_{2}/4H_{2}\simeq 0$( for the parameters considered in the article $m\simeq 10^{-7}$).
As one can see from this formula, the population imbalance is oscillating about non-zero average (MQST regime) with frequency given by
\begin{equation}
\omega=2g\left(\sqrt{1+\Delta^2}+\frac{(Z(0)\Delta-\sqrt{1-Z(0)^2})(2\Delta^2-1)}{2(1+\Delta^2)^{3/2}}\lambda\rho \right)
\end{equation}
 This two regimes  are shown  in Fig.\ref{graph liambda=0.1}.


\subsection{\protect\normalsize Population imbalance and oscillation frequencies   for the intermediate values of $\lambda \rho$  }
\begin{figure}
\begin{center}
{\includegraphics*[scale = 0.6]{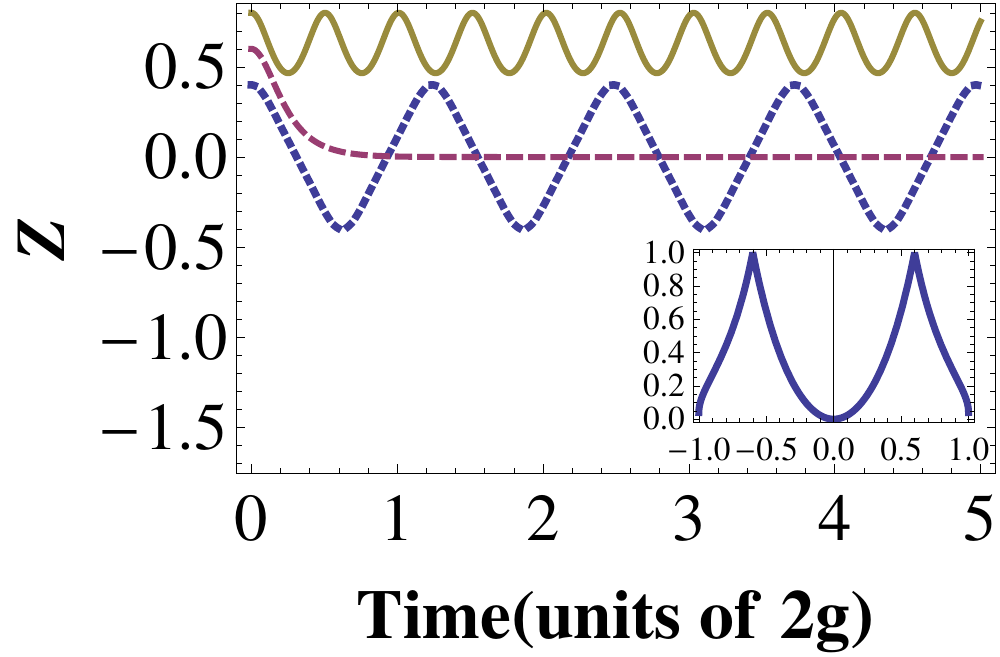}}
\end{center}
\caption{(Color
online). The population imbalance  in  two coupled rings for the intermediate value of $\lambda \rho$ and  $\Delta = 0$.
Dotted line,dashed line and solid line  respectively correspond to the cases $k<1, k=1, k>1$.
Inset shows dependence of the elliptic modulus $\tilde{k}$ $(\tilde{k}=k,\ \ for \ \  k<1;\ \ 1 \ \ for \ \ k=1 \ \ and \ \ 1/k,\ \ for \ \ k>1)$ from the value of $Z_0$.
Here $\lambda\rho = 10$, $\Theta_{0}=0$.}
\label{graph liambda=10}
\end{figure}

{I-C} $\Delta=0$-- The population imbalance can be expressed in terms of Jacobi functions  'cn' and 'dn' (Eq.(\ref{sol1})) and behavior of the solutions (which is summariZed on Fig.\ref{graph liambda=10})  is governed by elliptic modulus $k$ (Eq.(\ref{el mod})).

\begin{figure}
\begin{center}
{\includegraphics*[scale = 0.6]{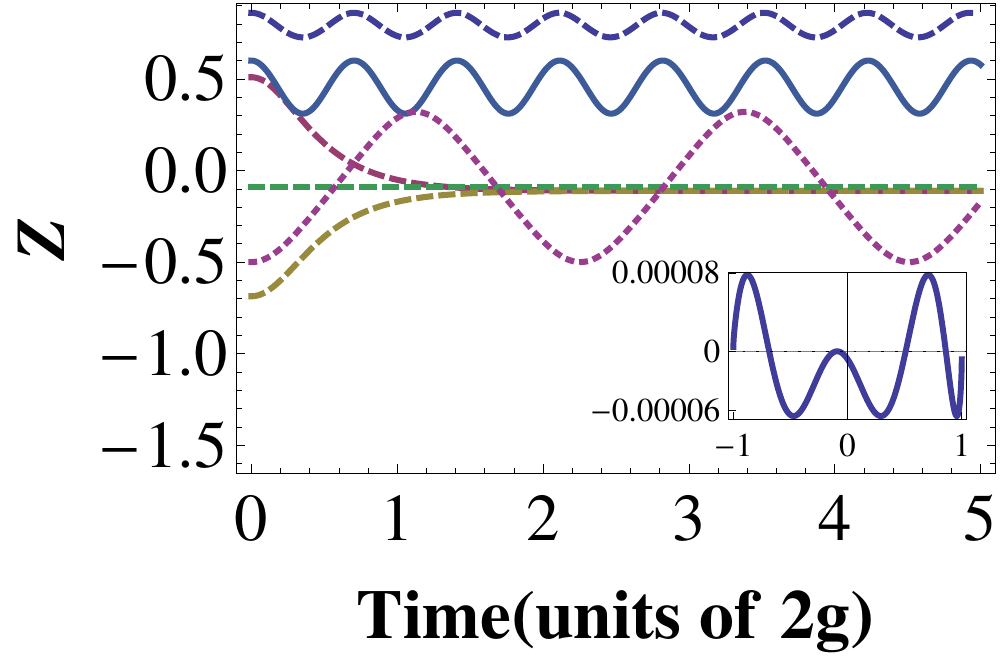}}
\end{center}
\caption{(Color
online). The population imbalance  in  two coupled rings for the intermediate value of $\lambda \rho$ and  $\Delta = 1$.
Dotted line, dashed line and solid line  respectively correspond to the cases $\delta<0, \delta=0, \delta>0$.
Inset shows dependence of determinant $\delta$ of characteristic cubic equation   from the value of $Z_0$.
Here $\lambda\rho = 10$, $\Theta_{0}=0$}
\label{graph1 liambda=10}
\end{figure}

{ II-C}   $\Delta\neq0$--  The population imbalance can be written in terms of Weierstrass elliptic function (Eq.(\ref{weier sol})) behavior of the solutions is governed by determinant $\delta$ (Eq.(\ref{discrim})) of the characteristic cubic equation.The dynamics for this subcase is given by Fig.\ref{graph1 liambda=10}.In  both cases oscillation periods can be expressed in terms of elliptic integral of the first kind (Eqs.(\ref{period1}) and (\ref{period2})). When $\delta=0$ oscillations are exponentially suppressed or there are sinusoidal oscillations depending on the relative sign between $g2$ and $g3$(Eqs.(\ref{sol-delta=0(1)}) and (\ref{sol-delta=0(2)})). Because it is  possible to express Weierstrass function through Jacobian functions 'sn' and 'cn' (Eqs.(\ref{sol-delta>0}) and (\ref{sol-delta<0})) we are coming to conclusion that in general population imbalance can be written in terms of Jacobian functions.\\

\subsection{\protect\normalsize Population imbalance and oscillation frequencies   for  $\lambda \rho\rightarrow\infty$  }

In analogy to a non-rigid pendulum, it is expected that in this regime, no oscillations occur and the population imbalance stays constant: the parameter $\lambda \rho$ is playing  the role of the moment of inertia and when it is very big it is not possible to force the pendulum to rotate  by providing it with finite amount of angular momentum (which is $Z_0$) or by acting on it with finite driving force $\Delta$. It is also possible to arrive at this conclusion from the analytical solutions shown in Appendix \ref{elliptic}.

I-D $\Delta =0$-- We see that $k\rightarrow\infty$ and $Z(\tilde {s})=C dn[(C\lambda \rho/k(\tilde {s}-\tilde {s}_{0}),0)=Z_0=const]$.

II-D
$\Delta \neq 0$-- In this case  $\delta=0$ and solution is given by Eq.(\ref{sol-delta=0(2)}). But because in this limit $f^{\prime}(Z_1)\rightarrow0$ and $Z_1=Z_0$ we conclude that $Z(\tilde {s})=Z_0=const$.

\subsection{\protect\normalsize  MQST and phase space diagrams}\label{sec:Allowed regions of oscillation}

In this section, it is shown how it is possible to find the values of the population imbalance which takes the system and the regions of MQST depending on $Z_0$.The phase diagrams in the $\Theta/ \pi,Z(\tilde{s})$ space for the cases of zero and non-zero values of the $\Delta$ are introduced as well.

\begin{figure}
\begin{center}
{\includegraphics*[scale = 0.6]{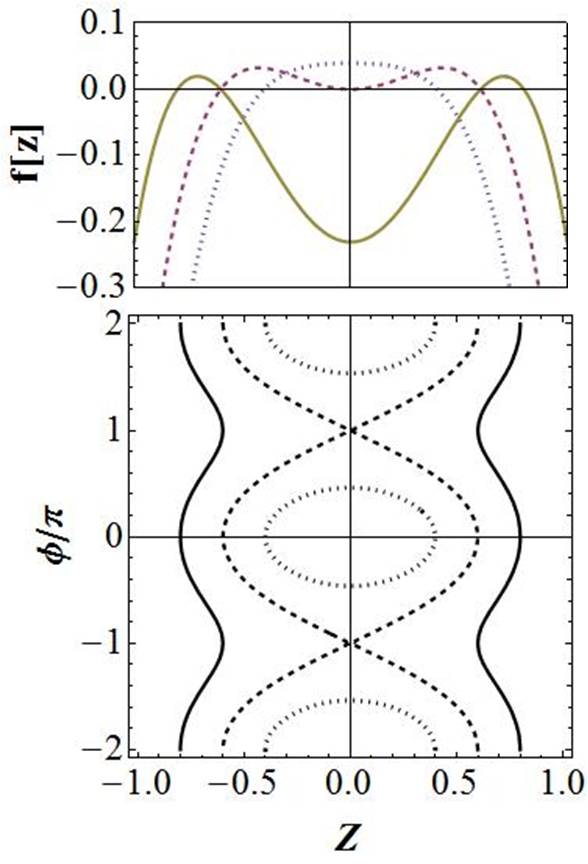}}
\end{center}
\caption{(Color
online). Dependence of the characteristic function $f(Z)$ on population imbalance(top) and phase space diagram(bottom) for the case $\Delta=0$. Oscillations can occur only in the regions where $f(Z)\geq0$. Dotted, dashed and solid lines respectively correspond to the values $Z_0=0.4,0.6,0.8$(for these values $k<1,k=1$ and $k>1$ respectively).Here $\Theta_0=0$ and $\lambda\rho = 10$.}
\label{graph phase space1}
\end{figure}

\begin{figure}
\begin{center}
{\includegraphics*[scale = 0.6]{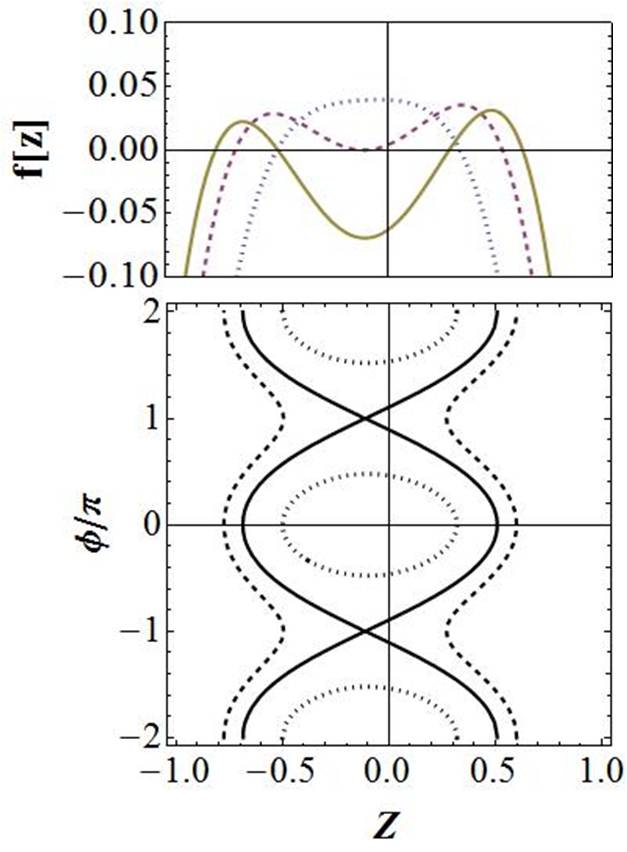}}
\end{center}
\caption{(Color
online). Dependence of the characteristic function $f(Z)$ on population imbalance(top) and phase space diagram(bottom) for the case $\Delta\neq0$. Oscillations can occur only in the regions where $f(Z)\geq0$. Dotted, dashed and solid lines respectively correspond to the values $Z_0=-0.5,0.509117,0.6$(for these values $\delta<0,\delta=0$ and $\delta>0$ respectively).Here $\Theta_0=0$ and $\lambda\rho = 10$.}
\label{graph phase space2}
\end{figure}

As it is seen from Eq.(\ref{Z solution}),  if the characteristic quartic equation (which is given be Eq.(\ref{quartic})) $f(Z)<0$, then the time takes  imaginary values. So we conclude that  allowed regions of $Z(\tilde{s})$ are given by the condition
\begin{equation}
f(Z)\geq0,
\label{f cond}
\end{equation}
Eq.(\ref{Z equation}), derived from the Gross-Pitaevskii Eqs.(\ref{dynamical_equation1}) and (\ref{dynamical_equation2}), can be written as an equation of motion of a classical particle with a coordinate $Z$, potential energy  $U(Z)$ and total energy $E$:
\begin{equation}
\dot{Z}^2(\tilde{s})+U(Z)=E,
\end{equation}
where the first term is playing the role of kinetic energy. The second term and the total energy are given by:
\begin{eqnarray}\nonumber
U(Z)&=&Z^2(\frac{{(\lambda \rho)}^2 Z^2}{4}+1+\Delta^2-H_0 \lambda \rho)+Z(\lambda \rho \Delta Z^2-2H_0 \Delta) \\
E&=&\dot{Z}^2(0)+U(Z(0))=1-H_{0}^{2}
\end{eqnarray}
Within the classical mechanics  analogy, $f(Z)$ plays  the role of $E-U$. The motion of the particle lies within the regions of the classical turning points in which the total energy equals to the potential energy. So the Eq.(\ref{f cond}) has a simple physical meaning: that classical particle can only move in the regions where total energy is equal or bigger  than potential energy.
From the upper graphs of the Figs.\ref{graph phase space1} and \ref{graph phase space2}, one can see that when all values of parameters are fixed,  and we start to change the value of $Z_0$ then the function $f(Z)$ changes from parabolic to double-well.

When $\Delta=0$ then for the parabolic potential $Z(\tilde{s})$ oscillates about an average of zero value,and when $Z(0)>Z_c$($(Z_c=0.6$ in our case) the particle is forced to oscillate about a non-zero average in one of the two wells as it is seen from Fig.\ref{graph phase space1} which is evidence of the MQST.
Indeed such phenomenon occurs in the system for  $Z(0)>Z_c$, where
\begin{widetext}
\begin{equation}
Z_c=\pm \sqrt{\frac{2}{\lambda \rho}-\frac{1+\cos{2\Theta_0}}{(\lambda \rho)^2}+\frac{\sqrt{(1+\cos{2\Theta_0})(\cos{2\Theta_0}+2(\lambda \rho-1)^2-1)}}{(\lambda \rho)^2}} \ \
\end{equation}
\end{widetext}
which reduces to
\begin{equation}
Z_c=\pm 2\frac{\sqrt{\lambda \rho-1} }{\lambda \rho}
\end{equation}
for $\sin{\Theta_0}=0$.
For the $Z_0=Z_c$, the particle moves from $Z_0$ to the point $Z=0$ where it stays for an infinite time because at this point $f^{\prime}(Z)=0$ which means there is no force  acting on the particle($U^{\prime}(Z)=0$).

When  $\Delta\neq0$, all the regimes are the same but with the difference that the external force $\Delta$ breaks the symmetry at the point $Z=0$, with the following two consequences: $(1)$:  $Z(\tilde{s})$ is always oscillating about a non-zero average, and $(2)$: for the critical value $Z_c $, the oscillations are damping to the non-zero value of population imbalance as one can see from Fig.\ref{graph phase space2}.So for the all values of the $Z_0$(excepting the values for which $\delta=0$) the effect of MQST occurs in the system.

The lower graphs of the Figs.\ref{graph phase space1} and \ref{graph phase space2} display corresponding phase versus population imbalance.
After expressing the phase using the population imbalance from Eq.(\ref{eff ham}), we get
\begin{equation}
\Theta=\arccos[\frac{H_{0}-\frac{\lambda \rho Z^2}{2}-\Delta Z}{\sqrt{1-Z^2}}]
\label{phase}
\end{equation}
From Fig.\ref{graph phase space1} it is seen that  when $Z<Z_c$, the phase diagram is a closed curve and the phase is oscillating about zero average value.  At the critical point, the phase diagram starts to split into the two open curves and for $Z>Z_c$ it consists of   two open curves and the phase takes  any value in the interval $[-\infty,\infty]$.In the case when $\Delta \neq 0$ the phase diagram is changing depending on $Z_0$ in a similar way but with the difference that the external force $\Delta$ is breaking the symmetry about the origin of $Z$ axis as demonstrated in Fig.\ref{graph phase space2}.
\section{Conclusions}\label{sec:conclusions}
In this paper,  we study the dynamics of a  physical system of two Bose-Einstein condensates, flowing in ring-shaped optical potentials, and mutually interacting through tunnel coupling.
The experimental setup, provided in Fig.\ref{double-ring potential}, allows us to tune the tunneling in an experimentally feasible way.
The system is  governed by a Bose-Hubbard two-leg ladder Hamiltonian, pierced by a synthetic magnetic field, effectively twisting the boundary conditions of the two rings.
We remark that the Galilean symmetry is broken in such system, along the ring-ring coupling direction.
In the weak ring-ring coupling regime,  the microscopic degrees of freedom (the phase slips between  adjacent wells along the rings) are  integrated out, leading to an effective action of a two level system  for the ring-ring phase slip. This implies that such physical system  provides indeed a qubit, analogous to the flux qubit built with superconductor Josephson junctions. It would be interesting to study the effective dynamics beyond the weak coupling approximation.

The real time evolution of  population imbalance and phase difference between the two flowing condensates  is analyzed, through two coupled  Gross-Pitaevskii equations obtained from the Bose-Hubbard ladder in the mean-field equations. The dynamics is thoroughly analyzed in the different regimes depending on the values of the s-wave scattering, tunneling rate, synthetic 'magnetic fluxes' and initial population imbalances  ($\lambda \rho$,$g$ ,$\Delta$ and $Z_0$ respectively).  It is evidenced how the macroscopic quantum self trapping  occurs in our system. It is shown that population imbalance takes values in only classically allowed regions of oscillation.

It would be interesting to describe our system including the quantum fluctuations around the mean field solution\cite{Maria Rey1, Maria Rey2}.

\acknowledgements
We thank  F. Auksztol, H. Crepaz,  R. Dumke, F. Hekking, A.J. Leggett, A. Minguzzi, and G. Rastelli  for discussions.

\begin{appendix}

\section{Peierls substitution for the Bose-Hubbard ladder model}
\label{peierls}
In this  appendix  we review the 'Peierls substitution'  in the Bose-Hubbard ladder, corresponding  to applying of two different 'magnetic fluxes'  to  the two-ring lattice system  (see Eq.(\ref{model})).

The hopping element $t$ can be expressed through Wannier functions $\phi(\textbf{x}-\textbf{R}_i)$ and single particle Hamiltonian $h_1$ in the subsequent form:
\begin{equation}
t_0=\int dx^3\phi^{*}(\textbf{x}-\textbf{R}_i)h_1\phi(\textbf{x}-\textbf{R}_{i+1}),
\label{tun el}
\end{equation}
The Wannier functions $\phi(\textbf{x}-\textbf{R}_i)$ are localized around $\textbf{R}_i$ lattice sites.
In the absence of 'electromagnetic fields' the  single particle Hamiltonian is given by
\begin{equation}
h_1=\frac{\textbf{p}^2}{2m}+V(\textbf{x}),
\end{equation}
where the first term is kinetic energy and second term is a one-body potential energy.
Once the synthetic gauge field $\textbf{A}(\textbf{x},t)$ is generated we can take it into account by substitution $\textbf{p} \rightarrow \textbf{p}-\textbf{A}(\textbf{x},t)$ in the single-particle Hamiltonian. We can rewrite  the hopping element  in the presence of the synthetic gauge field in the following form:
\begin{equation}
t=\int dx^3 \tilde{\phi}^{*}(\textbf{x}-\textbf{R}_i)h_1\tilde{\phi}(\textbf{x}-\textbf{R}_{i+1}),
\label{tun el1}
\end{equation}
where $\tilde{\phi}(\textbf{x}-\textbf{R}_i)=e^{-i \Lambda(\textbf{x},t)} \phi^{*}(\textbf{x}-\textbf{R}_i)$ with $\Lambda(\textbf{x},t)=\int_{\textbf{x}_0}^{\textbf{x}} \textbf{A}(\textbf{x},t)\textbf{dx}$,where $\textbf{x}_0$ is an arbitrary point. Assuming that $\textbf{A}(\textbf{x},t)$  is slowly varying function on an atomic scale
\begin{equation}
\tilde{\phi}(\textbf{x}-\textbf{R}_i)\approx e^{-i \Lambda(\textbf{R}_i,t)} \phi^{*}(\textbf{x}-\textbf{R}_i),
\label{approx}
\end{equation}
By substituting Eq.(\ref{approx}) in Eq.(\ref{tun el1}) we finally get
\begin{equation}
t=e^{i\Phi}t_0, \  \ \Phi =\int_{\textbf{R}_i}^{\textbf{R}_{i+1}} \textbf{A}(\textbf{x},t)\textbf{dx}
\end{equation}
The idea, that all the effect of electromagnetic field in the lattice can be absorbed in the hopping matrix element is called Peierls substitution.
We would like to emphasize, that the inter-ring hopping element $g$ is not affected by the Peierls substitution because  the synthetic gauge field is assumed to  have components longitudinal  to the rings only.

\section{Solution in terms of elliptic functions}
\label{elliptic}
As it was shown in Sect. \ref{model},  the dynamics of the population imbalance and the phase difference of the condensates in the two coupled rings is given by
\begin{equation}
\frac{\partial Z }{\partial \tilde {s}}=-\sqrt{1-Z^{2}}\sin{\Theta}
\label{dyn1}
\end{equation}
\begin{equation}
\frac{\partial \Theta }{\partial \tilde {s}
}=\Delta+\lambda \rho Z+\frac{Z}{\sqrt {1-Z^{2}}}\cos{\Theta}
\label{dyn2}
\end{equation}
In this appendix  we discuss the analytical  solutions of the equations above, for the two different cases: $\Delta=0$ and $\Delta\neq0$.

The   Eqs.(\ref{dyn1}) and (\ref{dyn2})  can be derived from the  Hamiltonian
\begin{equation}
H(Z(\tilde {s}),\Theta(\tilde {s}))=\frac{\lambda \rho Z^2}{2}+\Delta Z-\sqrt{1-Z^2}cos\Theta=H_0,
\label{Ham}
\end{equation}
where  $Z$ and $\Theta$ are  canonically  conjugate variables.
Indeed, $H(Z(t),\Theta(t))=H(Z(0),\Theta(0))=H_{0}$ because the  energy of   the system is conserved.
Combining Eqs.(\ref{dyn1}) and  (\ref{Ham}) $\Theta$ can be eliminated, obtaining
\begin{equation}
\dot{Z}^2+[\frac{\lambda \rho Z^2}{2}+\Delta Z-H_0]^2=1-Z^2,
\label{Z equation}
\end{equation}
that is solved  by  quadratures:
\begin{equation}
\frac{\lambda\varrho \tilde {s}}{2}=\int_{Z(0)}^{Z(\tilde {s})}\frac{dZ}{\sqrt{f(Z)}}
\label{Z solution}
\end{equation}
where $f(Z)$ is the  following quartic equation
\begin{equation}
f(Z)=\big(\frac{2}{\lambda \rho}\big)^2(1-Z^2)-\big[Z^2+\frac{2Z\Delta}{\lambda \rho}-\frac{2H_0}{\lambda \rho}\big]^2
\label{quartic}
\end{equation}

There are two different cases: $\Delta=0$ and $\Delta\neq 0$.

{ {I)}}  $\Delta=0.$ --
In this case the solution for the $Z(t)$ can be expressed in terms of  'cn' and 'dn' Jacobian elliptic functions as(\cite{Smerzi}):
\begin{eqnarray}
Z(\tilde {s})&=&C cn[(C\lambda \rho/k(\tilde {s}-\tilde {s}_{0}),k)] \ \ for \ \ 0<k<1 \nonumber \\
&=& C sech(C\lambda \rho(\tilde {s}-\tilde {s}_{0})), \ \ for \ \ k=1 \nonumber \\
&=&C dn[(C\lambda \rho/k(\tilde {s}-\tilde {s}_{0}),1/k)] \ \ for \ \ k>1; \label{sol1} \\
k&=&\big(\frac{C\lambda \rho}{\sqrt{2}\zeta(\lambda \rho)}\big)^2=\frac{1}{2}\big[1+\frac{(H_{0}\lambda \rho-1)}{(\lambda \rho)^2+1-2H_{0}\lambda \rho}\big],\label{el mod}
\end{eqnarray}
where
\begin{eqnarray}
C^2=\frac{2}{(\lambda \rho)^2}((H_{0}\lambda \rho-1)+\zeta^2),\nonumber \\
\alpha^2=\frac{2}{(\lambda \rho)^2}(\zeta^2-(H_{0}\lambda \rho-1)), \nonumber \\
\zeta^2(\lambda \rho)=2\sqrt{(\lambda \rho)^2+1-2H_{0}\lambda \rho},\label{paramet}
\end{eqnarray}
and $\tilde {s}_{0}$  fixing $Z(0)$. Jacobi functions are defined in terms of the incomplete elliptic integral of the first kind  $F(\phi,k)=\int_{0}^{\phi}d\theta(1-k \sin^2{\theta})^{-1/2}$ by the following expressions: $sn(u|k)=\sin{\phi},cn(u|k)=\cos{\phi}$ and $dn(u|k)=(1-k \sin^2{\phi})^{1/2}$~\cite{Abramowitz}. The Jacobian elliptic functions $sn(u|k)$, $cn(u|k)$ and $dn(u|k)$ are periodic in the argument $u$ with period $4K(k)$, $4K(k)$ and $2K(k)$,  respectively, where  $ K(k)=F(\pi/2,k)$ is the complete elliptic integral of the first kind. For small elliptic modulus $k\simeq0$  such functions  behave as trigonometric functions; for  $k\simeq1$ they behave as hyperbolic functions. Accordingly, the  character of the solution of Eqs.(\ref{dyn1}) and (\ref{dyn2}) can be   oscillatory or exponential,  depending  on $k$.
For  $k\ll1$, $cn(u|k)\approx\cos{u}+0.25 k (u-\sin{(2u)}/2)\sin{u}$ is almost sinusoidal and the population imbalance is oscillating around zero average value. When $k$ increases, the  oscillations  become non-sinusoidal and for $1-k\ll1$ the time evolution  is non-periodic: $cn(u|k)\approx\sec{u}-0.25 (1-k) (\sinh{(2u)}/2-u)\tanh{u}\sec{u}$. From the last expression we can see that at $k=1$, $cn(u|k)=\sec{u}$ so oscillations are exponentially suppressed and $Z(\tilde{s})$ taking $0$ asymptotic value. For the values of the $k>1$ such that $[1-1/k]\ll1$ and $Z(s)$ is still non-periodic and is given by: $dn(u|1/k)\approx\sec{u}+0.25 (1-1/k) (\sinh{(2u)}/2+u)\tanh{u} \sec{u}$.  Finally when $k\gg1$ then the behavior switches to   sinusoidal again,  but $Z(\tilde{s})$ does oscillates around  a non-zero average: $dn(u|1/k)\approx1-\sin^2{u}/2k$. This phenomenon accounts for the MQST.
%


The   periods of oscillations  in the regimes considered above result to be
\begin{eqnarray}\nonumber
\tau &=&\frac{4kK(k)}{C\lambda \rho} \ \ for  \ \ 0<k<1, \nonumber \\
&=& log{(4/\sqrt{1-k})} \ \ for \ \ k=1, \nonumber \\
&=& \frac{2K(1/k)}{C\lambda \rho} \ \ for \ \ k>1
\label{period1}
\end{eqnarray}
For $k\rightarrow1$ the period becomes infinite and diverging logarithmically.

{ {II)}}$\Delta \neq 0 $.--
In this case $Z(s)$ is expressed in terms of the Weierstrass elliptic function(\cite{Kenkre,Smerzi}):
\begin{equation}
Z(\tilde {s})=Z_1+\frac{f^{\prime}(Z_1)/4}{\varrho(\frac{\lambda \rho}{2}(\tilde {s}-\tilde {s}_{0});g_{2},g_3)-\frac{f^{\prime\prime}(Z_1)}{24}}
\label{weier sol}
\end{equation}
where $ f(Z) $ is given by an expression (\ref{quartic}), $Z_1$ is a root of quartic $f(Z)$ and $\tilde {s}_{0}=(2/\lambda \rho)\int_{Z_1}^{Z(0)}\frac{dZ^{\prime}}{\sqrt{f(Z^{\prime})}}$. For $\sin{\Theta_0}=0$ (which is the case discussed in the text),  $Z_1=Z_0$ and consequently $s_0=0$.
The Weierstrass elliptic function can be given as the inverse of an elliptic integral $\varrho(u;g_2,g_3)=y$, where
\begin{equation}
u=\int_y^\infty \frac{ds}{\sqrt{4s^3-g_2 s-g_3}} \; .
\end{equation}
The constants $g_2$ and $g_3$ are  the characteristic  invariants  of $\varrho$:
\begin{eqnarray}\nonumber
g_2 &=& -a_4-4a_1a_3+3a_2^2 \\
g_3 &=&-a_2 a_4+2a_1 a_2  a_3-a_2^3+a_3^2-a_1^2a_4,
\end{eqnarray}
where the coefficients $a_i$, where $i=1,..,4$, are given as
\begin{eqnarray}\nonumber
a_1&=&-\frac{\Delta}{\lambda \rho};a_2=\frac{2}{3(\lambda \rho)^2}(\lambda \rho H_0-(\Delta^2+1)) \\
a_3&=&\frac{2H_0\Delta}{(\lambda \rho)^2};a_4=\frac{4(1-H_0^2)}{(\lambda \rho)^2}
\end{eqnarray}
In the present case $\Delta \neq 0 $, the discriminant
\begin{equation}
\delta=g_2^3-27g_3^2
\label{discrim}
\end{equation}
 of the cubic $h(y)=4y^3-g_2y-g_3$ governs the behavior of the Weierstrass elliptic functions (we contrast with
 the case  $\Delta=0$, where  the dynamics is governed by the elliptic modulus $k$).
%

At first we consider the case $\delta=0$.

If $g_2<0$,$g_3>0$ then(\cite{Abramowitz})
\begin{equation}
Z(\tilde {s})=Z_1+\frac{f^{\prime}(Z_1)/4}{c+3c\sinh^{-2}{[\frac{\sqrt{3c}\lambda \rho}{2}(\tilde {s}-\tilde{s}_{0})]}-\frac{f^{\prime\prime}(Z_1)}{24}}
\label{sol-delta=0(1)}
\end{equation}
Namely, the  oscillations of $Z$ are exponentially suppressed  and the population imbalance decay (if $Z_0>0$) or saturate (if $Z_0<0$) to the asymptotic value given by
$Z(\tilde{s})=Z_1+\frac{f^{\prime}(Z_1)/4}{c-f^{\prime\prime}
 (Z_1)/24}$. \\
If $g_2>0$,$g_3>0$ then(\cite{Abramowitz})
\begin{equation}
Z(\tilde {s})=Z_1+\frac{f^{\prime}(Z_1)/4}{-c+3c\sin^{-2}{[\frac{\sqrt{3c}\lambda \rho}{2}(\tilde {s}-\tilde {s}_{0})]}-\frac{f^{\prime\prime}(Z_1)}{24}}
\label{sol-delta=0(2)}
\end{equation}
where $c=\sqrt{g_2/12}$.
We see that the population imbalance  oscillates around a non-zero average value $\overline{Z}\doteq Z_1+\frac{f^{\prime}(Z_1)/4}{2(2c-f^{\prime\prime}(Z_1)/24)}$,  with  frequency
$\omega=2g \sqrt{3c}\lambda\rho$.

We  express the Weierstrass function in terms of Jacobian elliptic functions. This leads to  significant  simplification  for the analysis of these regimes.

For $\delta>0$, it results
\begin{equation}
Z(\tilde {s})=Z_1+\frac{f^{\prime}(Z_1)/4}{e_3+\frac{e_1-e_3}{sn^{2}[\frac{\lambda \rho \sqrt{e_1-e_3}}{2}(\tilde {s}-\tilde {s}_{0}),k_1]}-\frac{f^{\prime\prime}(Z_1)}{24}}
\label{sol-delta>0}
\end{equation}
where $k_1=\frac{e_{2}-e_{3}}{e_{1}-e_{3}} $ and $e_{i}$ are  solutions of the cubic equation $h(y)=0$.
 In this case   the population imbalance oscillates  about the average value $\overline{Z}=Z_1+\frac{f^{\prime}(Z_1)/4}{2(e_1-f^{\prime\prime}(Z_1)/24)}$.

The asymptotics of the solution is extracted through:  $k\ll1$, $sn(u|k)\approx\sin{u}-0.25 k (u-\sin{(2u)}/2)\cos{u}$. When $k$ increases oscillations starting to become non-sinusoidal and when $1-k\ll1$ it becomes non-periodic and takes form: $cn(u|k)\approx\tanh{u}-0.25 (1-k) (\sinh{(2u)}/2-u)\sec^2{u}$.

For $\delta<0$ the following expression for $Z(s)$ is obtained:
\begin{equation}
Z(\tilde {s})=Z_1+\frac{f^{\prime}(Z_1)/4}{e_2+H_2\frac{1+cn[\lambda \rho \sqrt{H_2}(\tilde {s}-\tilde {s}_{0}),k_2]}{1-cn[\lambda \rho \sqrt{H_2}(\tilde {s}-\tilde {s}_{0}),k_2]}-\frac{f^{\prime\prime}(Z_1)}{24}},
\label{sol-delta<0}
\end{equation}
where $k_2=1/2-\frac{3e_{2}}{4H_{2}}$ and $H_{2}=\sqrt{3e_{2}^2-\frac{g_{2}}{4}}$.The asymptotical behavior of the function $cn(u|k)$ has been discussed in the previous subsection.
As it it seen from this expression $Z(\tilde {s})$ oscillates  about the average value ${\overline{Z}}=Z_1+\frac{f^{\prime}(Z_1)/4}{2(e_2-f^{\prime\prime}(Z_1)/24)}$.

The period of the oscillations of the $Z(\tilde{s})$ in this case is given by
\begin{eqnarray}
\tau &=&\frac{K(k_{1})}{\lambda \rho \sqrt{e_{1}-e_{3}}} \ \ for  \ \ \delta>0, \nonumber \\
&=& \frac{K(k_{2})}{\lambda \rho \sqrt{H_{2}}} \ \ for \ \ \delta<0. \label{period2}
\end{eqnarray}
The inter-ring  tunneling Josephson current is given by
\begin{equation}
I=\frac{\dot{Z}N_{T}}{2}=I_{0}\sqrt{1-Z^2}\sin{\Theta},
\end{equation}

\end{appendix}

\end{document}